\documentclass[12pt]{article}

\usepackage{amsmath}
\usepackage{graphicx}

\usepackage{natbib}
\usepackage{psfrag}
\usepackage{amssymb}

\def\be{\begin{equation}}
\def\ee{\end{equation}}
\def\bea{\begin{eqnarray}}
\def\eea{\end{eqnarray}}
\def\bml{\begin{mathletters}}
\def\eml{\end{mathletters}}

\def\no{\nonumber}

\begin{document}

\title{Loss of least-loaded class in asexual populations due to drift and epistasis}
\author{Kavita Jain \\\mbox{}\\ Theoretical Sciences Unit and 
  Evolutionary and Organismal Biology Unit,\\Jawaharlal Nehru Centre
  for Advanced Scientific Research, \\ Jakkur P.O., Bangalore 560064, India}

\maketitle

\newpage

\noindent
Running head: Loss of least-loaded class
\bigskip


\bigskip

\noindent
Corresponding author: \\
Kavita Jain, \\
Theoretical Sciences Unit and Evolutionary and Organismal Biology
Unit,\\
Jawaharlal Nehru Centre for Advanced Scientific Research, \\ 
Jakkur P.O., Bangalore 560064, India. \\
\texttt{jain@jncasr.ac.in}

\bigskip

\noindent
\textbf{Abstract:} We consider the dynamics of a non-recombining 
haploid population of finite 
size which accumulates deleterious mutations irreversibly. This ratchet
like process occurs at a
finite speed in the absence of epistasis, but it has been
suggested that synergistic epistasis can  halt the 
ratchet. Using a diffusion theory, we find explicit analytical 
expressions for the typical time between successive clicks of the
ratchet for both non-epistatic and epistatic fitness
functions. Our calculations show that the inter-click time  
is of a scaling form which in the absence of epistasis gives a
speed that is determined by size of the least-loaded class and the selection
coefficient. With synergistic interactions, the ratchet speed is found
to approach zero rapidly for arbitrary epistasis. Our
analytical results are in good agreement with the numerical simulations.

\newpage
In an asexual population of size $N$,  
even the fittest individuals can be lost by stochastic
fluctuations arising due to the finiteness of the population size. 
If the individual's genome is long enough that the back mutations can be ignored
and recombination is absent, the minimum number of deleterious 
mutations (least-loaded class) in a finite population 
increases irreversibly \citep{Muller:1964, Felsenstein:1974}. 
For this reason, this process has been invoked as a potential cause
for the evolution of sex and recombination
\citep{Judson:1996,Hurst:1996,Barton:1998},  
degeneration of 
nonrecombining parts like $Y$ chromosome \citep{Charlesworth:1978} and 
mitochondrial DNA \citep{Loewe:2006} of sexually reproducing organisms
and extinction of obligately asexual populations by 
mutational meltdown \citep{Gabriel:1993}.

Due to the irreversible accumulation of deleterious mutations, the
process described above acts like a ratchet each click of which
corresponds to the loss of the least-loaded class. In the simplest
model known as the Muller's ratchet, the ratchet
clicks at a constant rate which depends on the population size $N$,
mutation rate $U$ and selection coefficient $s$. 
The ratchet speed is also known to depend on other biologically
relevant factors such as 
recombination rate \citep{Bell:1988,Charlesworth:1993}, epistatic interactions
\citep{Kondrashov:1994,Butcher:1995,Schultz:1997},  
fraction and selection coefficient of favorable mutations
\citep{Woodcock:1996,Bachtrog:2004} and spatial
structure of the population \citep{Combadao:2007}. Although much
numerical data for the ratchet speed is available, very few 
analytical results are known.

As it is desirable to stop or at least slow down the ratchet, several
mechanisms with this objective have been proposed
\citep{Bell:1988,Wagner:1990,Charlesworth:1993}. One such proposal
is to include  epistatic interactions in the genome fitness 
\citep{Charlesworth:1993,Kondrashov:1994}. It has been observed 
experimentally that the gene loci do not always contribute  
independently to the genome fitness \citep{Wolf:2000,Visser:2007} and
the effect of two deleterious
mutations can be better (antagonistic) or worse (synergistic) 
than were they to act independently. 
For Muller's ratchet operating under epistatic selection,  
 it was noted using numerical simulations that  ``sufficiently strong
 synergistic 
epistasis  can effectively halt the action of Muller's ratchet''
\citep{Kondrashov:1994}. However it was not shown how the ratchet
speed approaches
zero asymptotically and  how much epistatic interaction 
is required to halt the ratchet.

In this article, besides the classic Muller's ratchet 
 that assumes haploid asexual population evolving on a fitness
landscape in which each gene locus contributes independently to the
fitness of the genome, we also study Kondrashov's model which
considers fitness functions with epistatic interactions. We 
assume that an individual with $k$ mutations has a fitness 
\be
W(k)=(1-s)^{k^\alpha}
\label{fitness}
\ee
where $s$ is the selection coefficient.
 For $\alpha =1$, the epistatic interactions are
absent while $\alpha > 1$ corresponds to synergistically epistatic
fitness. Our 
main purpose is to obtain explicit analytical expressions for 
the typical time ${\overline T}_J$ elapsed between the $(J-1)$th and
$J$th click of
the ratchet in these models. If the time spent between any two
successive clicks is found to be constant then the ratchet turns 
with a finite speed $1/{\overline T}$, while it is said to be halted at large
times if the inter-click time increases with the
number of accumulated mutations. 

In the past, 
Muller's ratchet has been investigated using a diffusion
approximation which assumes that the population $n_0$ of 
the least loaded class is large and applies to slowly clicking
ratchet \citep{Stephan:1993,Gordo:2000a,Stephan:2002}. The opposite
 situation of small $n_0$ and high ratchet rate has been described by
 a  moments method 
 \citep{Gabriel:1993,Gessler:1995,Higgs:1995,Bennett:1997,Rouzine:2003}. In 
this article, we adopt the first method that 
works in the parameter range for which $n_0 \gg 1$ and the ratchet
clicks are slow enough that 
the population can equilibrate between successive clicks. This
requires the knowledge of the steady state properties of an infinitely
large population which are known
exactly for the Muller's ratchet \citep{Kimura:1966,Higgs:1994} but
have been studied numerically for
epistatic case \citep{Kondrashov:1994}. After defining the models in
the following section, we solve the deterministic quasispecies
equation in steady state 
for $\alpha > 1$ and show that the population frequency of the class
with minimum number $J$ of mutations increases with $J$. These
deterministic results are then 
used to find an expression for the typical time ${\overline T}_J$ in
terms of a double integral over the
frequency of the least-loaded class which 
have been evaluated numerically for $\alpha=1$
\citep{Stephan:1993,Gordo:2000a}. Here we estimate
these integrals analytically and find that for a broad range of parameters, the
average inter-click time is of a scaling form (\ref{scaling}) for any 
$\alpha \geq 1$. For 
$\alpha=1$, it is shown that the ratchet speed is a function of the
number $n_0= N e^{-U/s}$ in the least-loaded class and the selection
coefficient $s$, and not $n_0$ alone as assumed in previous studies 
\citep{Haigh:1978}. 
With epistatic interactions, the time ${\overline T}_J$
is found to increase faster than any power law with $J$ for any
$\alpha > 1$. Thus, an arbitrarily small amount of epistasis is
sufficient to halt the ratchet with the ratchet speed approaching zero
as $\sim 1/t$ for large time $t$. 

\bigskip
\centerline{MODELS}
\bigskip

We consider a haploid asexual population evolving via 
mutation-selection dynamics starting with an initial condition in
which all the individuals in the population have zero mutations. The
genome length is
assumed to be infinite so that back mutations can be ignored. If the
population has a finite size $N$, it evolves stochastically following the
discrete time Wright-Fisher dynamics. An offspring in generation $t+1$
chooses a parent in the previous generation with a probability
proportional to the fitness of the parent. Then the probability $P(n)$
that a parent $p$ carrying $k$ mutations and with fitness $W(k;p)$ 
has $n$ descendants in one generation is given by
\be
P(n)= {N \choose n} \left( \frac{W(k;p)}{N {\langle W \rangle}}
\right)^n \left(1-\frac{W(k;p)}{N {\langle W \rangle}} \right)^{N-n}
\label{binomial}
\ee
where ${\langle W \rangle}=\sum_{k=0}^{\infty} W(k) X(k,t)$ is the
average fitness of the finite population in generation $t$. Here we
have defined $X(k,t)$ as the fraction of population with $k$ mutations 
 in a single sampling of Wright-Fisher process. From the above
 equation, it follows that the average number of offspring produced in
 one generation is proportional
 to the parent's fitness and the relative variance in offspring number
 decays as $1/N$. This fact will be useful in defining the diffusion
 coefficient (\ref{D2}) within diffusion approximation discussed in later
 section. 
Following replication, mutations are introduced where the number of
new mutations is a random variable chosen from 
a Poisson distribution with mean $U$.  In the
simulations, the above process was implemented but the 
order of mutation and selection was reversed. An
individual picked randomly from the population
at time $t$ was first mutated and the resulting mutant was allowed to
survive at $t+1$ with a probability equal to its fitness. 
This process was repeated until the generation $t+1$ has
$N$ members and the population fraction $X(k,t+1)$ was recorded.  
It is useful to
define $X_J(k,t)=X(J+k,t)$ where $J$ is the minimum
number of mutations in the population at time $t$ so that 
$X_J(k,t)=0$ for $k < 0$. If $X_J(0,\tau_J)$ becomes zero, 
the least-loaded class $J$ is lost 
and the ratchet has clicked at time $\tau_J$.

The ratchet effect due to which the least-loaded class is lost 
is essentially a stochastic problem arising due to
the finite number $N$ of individuals in the population. However as we
shall describe later, the population fluctuates close to the
deterministic frequency between two clicks of the ratchet. 
For this reason, we also study the problem of  
an infinite population for which the 
density fluctuations vanish and the average population fraction 
${\cal X}(k,t)$ with $k$ mutations at time $t$ obeys a deterministic
quasispecies equation \citep{Eigen:1971,Jain:2007b}. Similar to the finite population problem, we 
define ${\cal X}_J(k,t)={\cal X}(J+k,t)$ where 
${\cal X}_J(k,t)=0$ for $k < 0$. Then neglecting the back mutations for
a genome of infinite length \citep{Higgs:1994}, 
the fraction ${\cal X}_J(k,t)$ evolves according to the following
difference equation, 
\be
{\cal X}_J(j,t+1)=\frac{1}{{\cal W}_J(t)} \sum_{k=0}^{j} e^{-U}
\frac{U^{j-k}}{(j-k)!} W(J+k) {\cal X}_J(k,t)~. 
\label{qsteqn}
\ee
In this equation, the population fraction with $k$ mutations replicates with
fitness $W(k)$ and accumulates further mutations which are Poisson
distributed with a mean $U$. The average fitness 
${\cal W}_J(t)= \sum_{k=0}^{\infty} W(J+k) {\cal X}_J(k,t)$ in the
denominator ensures that the number density is conserved.

\bigskip
\centerline{STEADY STATE OF THE QUASISPECIES MODEL}
\bigskip

In this section, we calculate the steady state population frequency 
${\cal X}_J(k)$ in the error class $J+k$. 
Unlike for the multiplicative fitness case, 
the frequency ${\cal X}_J(k)$ depends on $J$ 
for the epistatic fitness function \citep{Haigh:1978}. In
particular, the fraction ${\cal X}_J(0)$ (later abbreviated as ${\cal
  X}_J$) in the least-loaded class is expected to increase with $J$ 
for $\alpha > 1$ and decrease for $\alpha < 1$. 
This  can be explained by a simple 
argument which has also been used to understand the error threshold phenomenon 
\citep{Eigen:1971,Jain:2007b} in which the fittest genomic sequence can
get lost beyond a critical mutation rate 
in populations evolving on epistatic fitness landscapes
\citep{Wiehe:1997}. Consider the ratio 
$W(J+k)/W(J) \sim (1-s)^{\alpha k J ^{\alpha-1}}$ for $J \gg k$. For
synergistic  interactions, 
the error class $J+k$ in the neighborhood of the least-loaded class 
has a fitness
much worse than the fitness of class $J$ rendering selection effective in
localising population in the class with $J$ mutations. With increasing $J$, the
selection pressure increases further. Thus we may expect
the population frequency ${\cal X}_J(k)$ to peak around $J$ and ${\cal X}_J$ to
increase with $J$. On the other hand, in case of antagonistic
epistasis ($\alpha < 1$),
the fitness landscape is nearly neutral at large $J$ so that the
least-loaded sequence 
can be lost even in the deterministic limit (finite error
threshold) \citep{Wiehe:1997}.

In the steady state, the quasispecies equation (\ref{qsteqn}) reduces to 
\be 
{\cal X}_J(j)=\frac{1}{{\cal W}_J} \sum_{k=0}^{j} e^{-U}
\frac{U^{j-k}}{(j-k)!} W(J+k) {\cal X}_J(k) 
\label{qseqn}
\ee
where ${\cal W}_J$ is the average fitness in the steady state when 
the least-loaded class is $J$. 
The equation for $j=0$ immediately shows that \citep{Kimura:1966,Haigh:1978} 
\be
{\cal W}_J= W(J) e^{-U} ~. \label{avgfit}
\ee
For $j=1$ in (\ref{qseqn}), we have
\be
{\cal X}_J(1)=\frac{U W(J) {\cal X}_J}{W(J)-W(J+1)} \no~.
\ee
Plugging this expression in the equation for $j=2$, after some algebra 
we obtain
\be
{\cal X}_J(2)=\frac{U^2 W(J)  {\cal X}_J}{W(J)-W(J+2)}
\left[\frac{1}{2}+\frac{W(J+1)}{W(J)-W(J+1)}\right] \no~.
\ee
Similarly, the fraction in the error class $J+3$ is 
\bea
{\cal X}_J(3) &=&\frac{U^3 W(J) {\cal X}_J}{W(J)-W(J+3)}
\Big[\frac{1}{3!} +\frac{W(J+1)}{2! (W(J)-W(J+1))} {} \no \\ 
&+&  \frac{W(J+2)}{W(J)-W(J+2)} \left(\frac{1}{2!}+\frac{W(J+1)}{W(J)-W(J+1)} \right)\Big] \no~.
\eea
From the expressions for ${\cal X}_J(k)$ for $k=2, 3$ shown above,
it is clear that in the weak selection limit $s \to 0$, the leading
order contribution to ${\cal X}_J(k)$ 
comes from  the last term. In general, we can write 
\bea
{\cal X}_J(k) &\approx& \frac{U^k W(J) {\cal X}_J}{W(J+k)} \prod_{m=1}^{k}
\frac{W(J+m)}{W(J)-W(J+m)} \no \\
&\approx& \left(\frac{U}{s} \right)^k {\cal X}_J \prod_{m=1}^k \frac{1}{(J+m)^\alpha-J^\alpha}
\label{qssol}
\eea
where the population ${\cal X}_J$  in the
least-loaded class is determined using the normalisation condition 
$\sum_{k=0}^{\infty} {\cal X}_J(k)=1$.

\begin{figure}
\centerline{\includegraphics[width=0.42 \linewidth,angle=270]{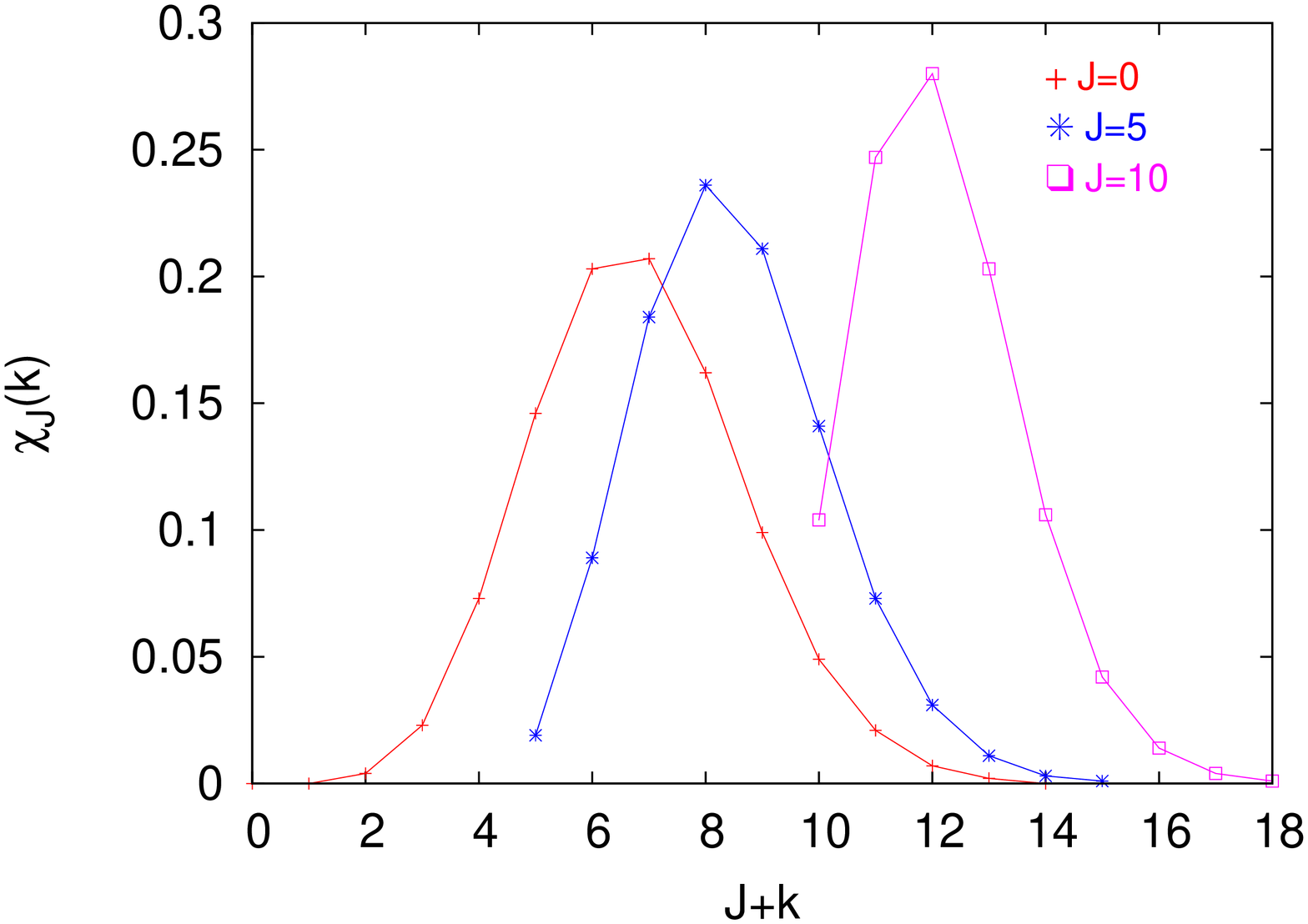}
\includegraphics[width=0.42 \linewidth,angle=270]{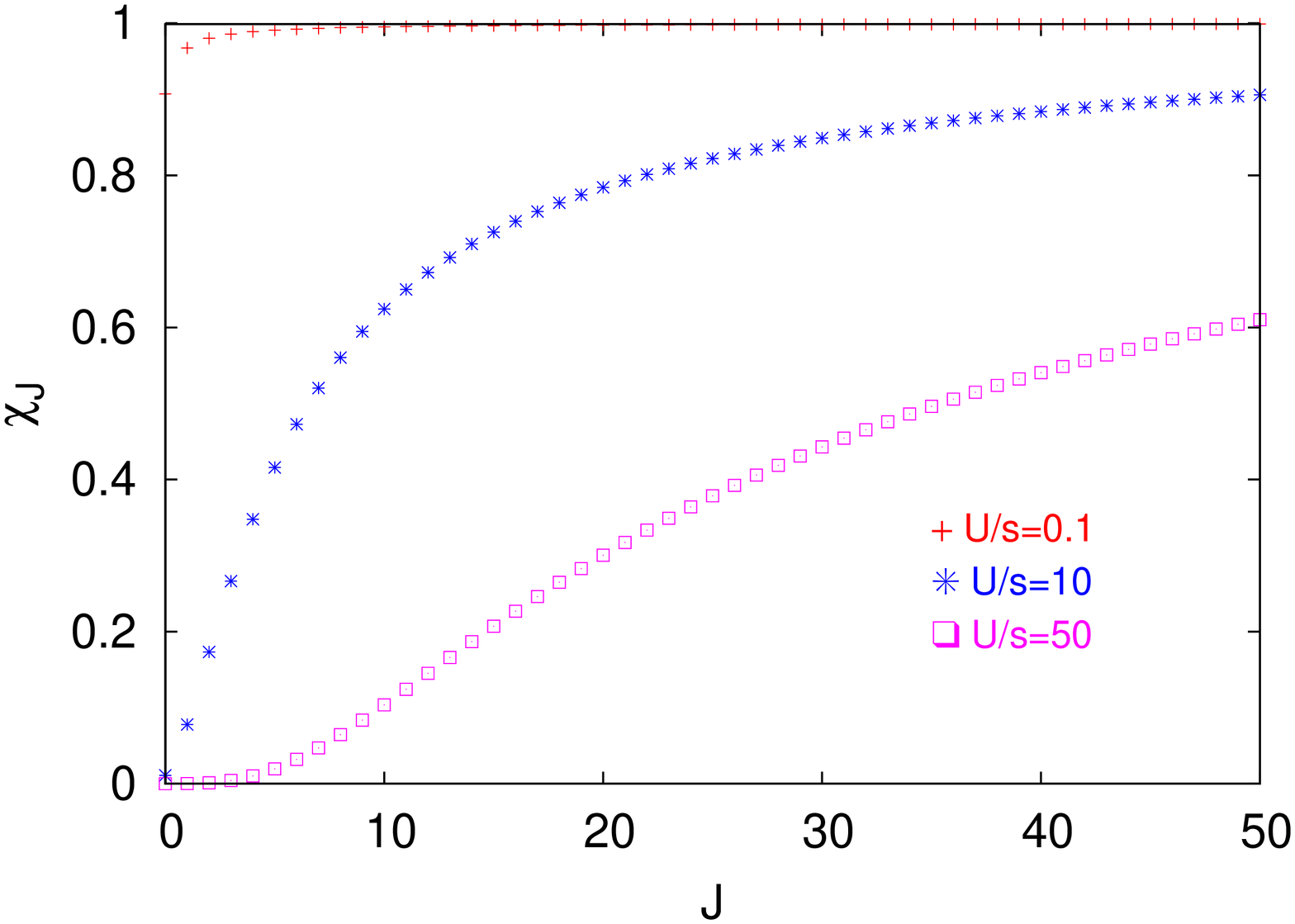}}
\caption{Steady state of the quasispecies model on synergistic
  fitness landscape with $\alpha=2$: (a) Fraction ${\cal X}_J(k)$ as a 
function of $J+k$ for $U/s=50$ given by (\ref{synform1}) (b)
Fraction ${\cal X}_J$ of the least-loaded class $J$ calculated using (\ref{synform2}).}
 \label{synfig}
\end{figure}

Using the preceding equation for the multiplicative fitness 
function, we obtain the well known result \citep{Kimura:1966}
\be
{\cal X}_J(k)=\left(\frac{U}{s}\right)^k\frac{{\cal X}_J}{k!}=
\left(\frac{U}{s}\right)^k\frac{ e^{-U/s}}{k!}~.
\ee
The fraction ${\cal X}_J(k)$ for all $k$ is seen to be independent of $J$ \citep{Haigh:1978}. 
For synergistically epistatic  fitness landscape with $\alpha=2$ in
fitness function (\ref{fitness}), we have 
\be
{\cal X}_J(k) \approx \left(\frac{U}{s}\right)^k {\cal X}_J
\prod_{m=1}^k \frac{1}{2 J m +m^2}=\left(\frac{U}{s}\right)^k \frac{(2
  J)!{\cal X}_J}{k ! (2 J+k)!}~.  
\ee
On summing both sides over $k$, we find 
\be
{\cal X}_J^{-1}=(2J)! \left(\frac{U}{s}\right)^{-J} I_{2 J} \left( 2 \sqrt{\frac{U}{s}} \right)
\label{synform2}
\ee
where $I_n(z)$ is the modified Bessel function of the first kind
\citep{Abramowitz:1964}. The fraction ${\cal X}_J(k)$ with $J+k$
mutations is then given by
\be
{\cal X}_J(k)=\left(\frac{U}{s}\right)^{J+k} \frac{1}{k ! (2 J+k)!} \frac{1}{I_{2 J} \left( 2 \sqrt{\frac{U}{s}} \right)}
\label{synform1}
\ee
and is plotted in Fig.~\ref{synfig}a as function of $J+k$ at a fixed 
$U/s$. 
From the above equation, we find that for a given $J$, the fraction 
${\cal X}_J(k)$ is centred about 
$k^*_2=\sqrt{J^2+J_2^2}-J$ where $J_2=\sqrt{U/s}$. For $J \ll J_2$, 
the distribution ${\cal X}_J(k)$ peaks about $k_2^* \approx J_2$ while for $J \gg J_2$, it is
maximised at $J_2^2/(2 J)$ (see Fig.~\ref{synfig}a). 
Thus for large $J$, as argued at the beginning of this
section, the distribution ${\cal X}_J(k)$ 
localises close to $k=0$. 
The behavior of the least-loaded fraction ${\cal X}_J$ shown in
Fig.~\ref{synfig}b also depends on $J_2$. For large $J_2$ (i.e. weak
selection), the fraction ${\cal X}_J$ increases towards unity slower
than for small $J_2$. 
Using the asymptotic expansion of the Bessel function $I_n(z)$ for
large orders \citep{Abramowitz:1964} in (\ref{synform2}), we have 
\be
{\cal X}_J^{-1} \sim \frac{(2J)!}{\sqrt{4 \pi J}}  \left(\frac{U}{s}\right)^{-J}  \frac{e^{2 J \sqrt{1+y^2}}}{(1+y^2)^{1/4}} \left(\frac{y}{1+\sqrt{1+y^2}} \right)^{2 J}
\ee
where we have defined $y=J_2/J$. For $J \gg J_2$, the above expression
can be simplified to give 
${\cal X}_J \approx {\rm exp}(-U/(2 s J))$ which asymptotically
approaches unity. Thus with increasing $J$, most of the population tends to 
stay in the least-loaded class. 

For arbitrary $\alpha > 1$, it does not seem possible to obtain explicit
expression for ${\cal X}_J(k)$. 
However using the insights obtained from $\alpha=2$ case, we can find 
${\cal X}_J$ for large $J$.   
We expect that for any $\alpha > 1$, there exists a least-loaded class $J_\alpha$ such that the population frequency  ${\cal X}_J(k)$  with  $J \gg J_\alpha$ is  nonzero for $k \ll J$. 
 In such a case, the denominator under the product sign in (\ref{qssol}) can be expanded for $m \ll J$ to leading orders and yield 
\be
{\cal X}_J(k) \approx \left(\frac{U}{\alpha s J^{\alpha-1}} \right)^k \frac{{\cal
    X}_J}{k !}~,~J \gg J_\alpha~.
\ee
As ${\cal X}_J(k)$ decays fast with $k$, we can sum over both sides of
the above solution to obtain
\be
{\cal X}_J \approx {\rm exp} \left[ -U/(\alpha s J^{\alpha-1})
  \right]~,~ J \gg J_\alpha~.
\label{synXJ}
\ee
This expression matches the exact results for $\alpha=1$ and $2$
discussed above. The product in (\ref{qssol}) seems hard to calculate for 
$J \ll J_\alpha$. But for $J=0$, we immediately have
\be
{\cal X}_{0}(k) \approx \left(\frac{U}{s} \right)^k \frac{{\cal
    X}_0}{k!^\alpha}
\ee
which peaks at $k_\alpha^*$ equal to $(U/s)^{1/\alpha}$. By analogy with the $\alpha=2$ case, this suggests $J_\alpha=(U/s)^{1/\alpha}$.

\bigskip
\centerline{TIME BETWEEN SUCCESSIVE CLICKS OF THE RATCHET}
\bigskip

\begin{figure}
\includegraphics[width=0.73 \linewidth,angle=270]{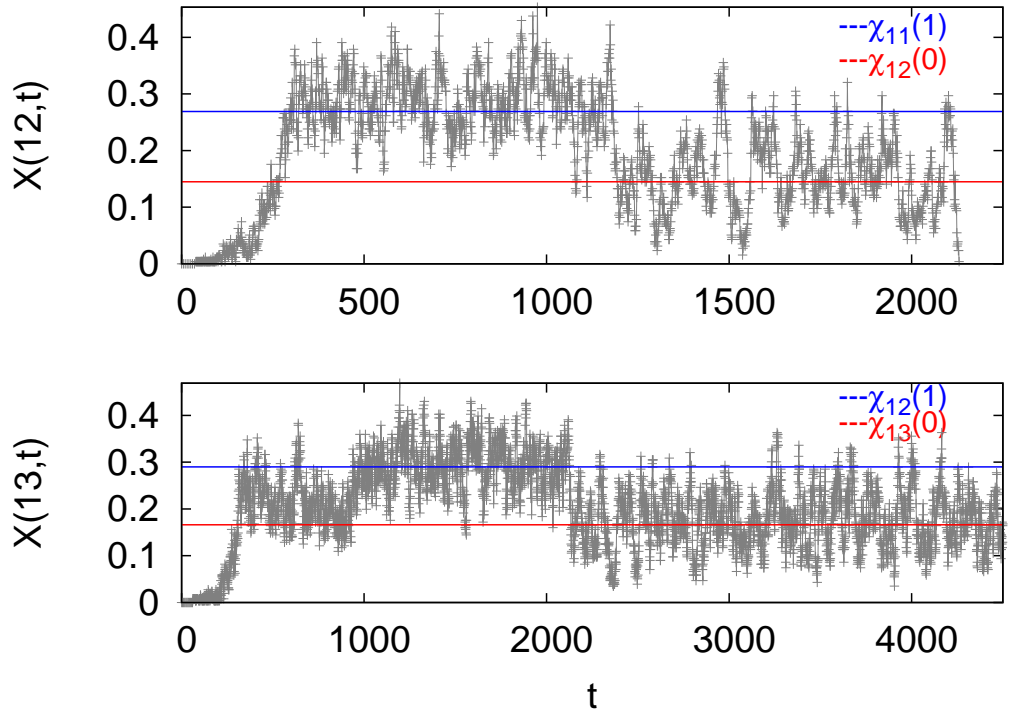}
\caption{Time evolution of the fraction $X(k,t)$ for $k=12$ and $13$ to
  illustrate the advance of the ratchet on epistatic fitness landscapes. Here 
$\alpha=2, N=256,  U=0.25$ and $s=0.005$. }
\label{synrun}
\end{figure}

In this section, we first describe the process by which ratchet clicks
and then calculate an expression for the typical time between successive
clicks using a diffusion theory \citep{Ewens:1979}.  
Let $X(k,t)$ denote
the fraction of population with $k$ mutations at time $t$ in a
single realization of the Wright-Fisher process. Figure~\ref{synrun}
shows the time evolution of $X(k,t)$ for $k=12$ and $13$ starting from
an initial condition in which all the $N$ individuals carry zero mutations. 
As Fig.~\ref{synrun}a illustrates, the fraction $X(12,t)$ increases
from zero to a steady state fraction about which it 
fluctuates until a time $\tau_{11}=1194$ after which it
relaxes to another steady state before finally dropping to zero at 
$\tau_{12}=2130$ due to stochastic fluctuations. At $t=\tau_{12}$, the
ratchet is said to have clicked as the least-loaded class with $12$
mutations gets irreversibly lost  
and the class with $13$ mutations shown in  Fig.~\ref{synrun}b becomes
the new least-loaded class which itself gets lost at $\tau_{13}=5235$. 
Since the ratchet is clicking at a slow rate, $X(k,t)$ 
has an opportunity to equilibrate. As 
Fig.~\ref{synrun}b shows, soon after time 
$\tau_{12}$, the fraction $X(13,t)$ fluctuates about a steady
state fraction which is close to the 
deterministic frequency ${\cal X}_{13}$ given by
(\ref{synform2}). Similarly, $X(12,t)$ in Fig.~\ref{synrun}a 
oscillates about ${\cal  X}_{12}$ after the $11$th error class is lost
until time $\tau_{12}$. As a click of the ratchet is signaled by the
change in the average value of $X(k,t)$, it follows that the $11$th
class is lost at $\tau_{11}$. 
For $ \tau_{11} < t < \tau_{12}$ as there are at least $12$ mutations
in the population,  the fraction 
$X(13,t)$ equilibrates about the frequency ${\cal X}_{12}(1)$. Thus
 the population fraction $X(k,t)$ for fixed
$k$  passes  through a series of steady states with
frequency ${\cal X}_J(k-J), J \leq k$ before reaching the final absorbing state 
$X(k,\tau_k)=0$. Note that this description of the mechanism by which the
ratchet clicks assumes that the population $N {\cal X}_J$ in the
currently least-loaded class $J$ far exceeds one and thus has a chance
to attain equilibrium before the next click.

We will use the diffusion approximation proposed in 
\citet{Stephan:1993} to find the average inter-click time 
$\overline{T}_J= \overline{\tau_J-\tau_{J-1}}$ between
the $(J-1)$th and $J$th click of the ratchet where $\overline{...}$
stands for averaging over stochastic histories. 
Let the random variable $X_J \in [0,1]$ denote the 
population fraction in the least-loaded class $J$. If $X_J=0$ at time $t$,
the current least-loaded class $J$ is lost forever and the ratchet is said to
have clicked at $t$. We are interested in calculating
the average time ${\overline T}_J$ required to reach the absorbing state
$X_J=0$ starting from $X'_J$ at $t'$. 
The probability distribution $P(X_J,t|X_J',t')$ obeys the following 
backward Fokker-Planck equation \citep{Risken:1996}, 
\bea
-\frac{\partial P(X_J,t|X_J',t')}{\partial t'}&=& D_1(X_J',t')
 \frac{\partial P(X_J,t|X_J',t')}{\partial X_J'} {} \no \\
&+&\frac{D_2(X_J',t')}
{2} \frac{\partial^2 P(X_J,t|X_J',t')}{\partial X_J'^2} \label{fokker}
\eea
where 
\bea
D_n(X_J',t') &=&{\rm  Lt}_{\tau  \to 0} \int_0^1 dX_J\frac{(X_J-X_J')^n
P(X_J,t'+\tau|X_J',t')}{\tau} \no \\
&=& {\rm  Lt}_{\tau  \to 0} 
\frac{\overline{\left[X_J(t'+\tau)-X_J(t') \right]^n}}{\tau}~.
\label{Dn}
\eea
As  the coefficients $D_n$ in (\ref{fokker}) 
are independent of $t'$ (see below), the average inter-click time 
${\overline T}_J$ defined as 
\be
{\overline T}(X_J')=\int_{-\infty}^0 dt'~ (-t') P(0,0|X_J',t')
\ee
obeys the following ordinary differential equation, 
\be
-1= D_1(X_J') \frac{d {\overline T}}{d X_J'}+\frac{D_2(X_J')}{2} \frac{d^2
  {\overline T}}{d X_J'^2}~.
\label{ode}
\ee
Since $X_J'=0$ is an absorbing state, the solution to the above equation 
is subjected to the boundary condition ${\overline
  T}(0)=0$. Furthermore as the population in the $J$th class 
equilibrates about the mean ${\cal X}_J$ after the $(J-1)$th click, we
can choose the initial distribution of random variable $X_J'$ 
to be $\delta(X_J'-{\cal X}_J)$.  
Then the time ${\overline T}_{J}$ during
which $J$ is the least loaded class obtained by solving
(\ref{ode}) is given by \citep{Ewens:1979} 
\be
{\overline T}_{J}=2 \int_0^{{\cal X}_J} dY ~ \psi(Y) \int_Y^1
\frac{dZ}{\psi(Z) D_2(Z)}
\label{TJ}
\ee
where $\psi(Y)={\rm exp}\left[-2 \int^Y dX~ D_1(X)/D_2(X) \right]$.

We will now determine the coefficients $D_1$ and $D_2$. The drift coefficient $D_1$ defined in (\ref{Dn}) measures the change
in the average fraction of the least-loaded class over a generation. 
As the population is in local equilibrium, this can be
determined using the quasispecies equation (\ref{qsteqn}) for $j=0$. 
Thus the drift coefficient is given by \citep{Stephan:1993,Gordo:2000a}
\bea
D_1(X_J={\cal X}_{J}(0,t))&=&{\cal X}_{J}(0,t+1)-{\cal X}_{J}(0,t) \no \\
&=& X_J \frac{e^{-U} W(J)-{\cal W}_J(t)}{{\cal W}_J(t)} \label{D1full}~.
\eea
As expected, $D_1$ vanishes when the population is either in the steady state
(see (\ref{avgfit})) or in the absorbing state ($X_J=0$).  
The equation (\ref{D1full}) for $D_1$ does not close in $X_J$ but one
can obtain an approximate expression for $D_1$ using the linear
response theory \citep{Risken:1996}. As $D_1$ is proportional to the
deviation from a steady state quantity, we
can write \citep{Stephan:1993}
\be
D_1 \propto e^{-U} W(J)-{\cal W}_J(t)= C \left(1- \frac{X_J}{{\cal X}_{J}}
\right) \label{dev}
\ee
where $C$ is a constant. Thus the drift coefficient can be written in
terms of $X_J$ as 
\be
D_1 \approx C X_J \left(1- \frac{X_J}{{\cal X}_{J}}\right)~.
\label{D1new}
\ee
As Fig.~\ref{synrun} shows, when the ratchet
clicks ($X_J=0$) and the $J$th class is lost, the
population quickly relaxes to the equilibrium frequency of the
$(J+1)$th class so that the
deviation in the fitness $C \sim {\cal W}_J-{\cal W}_{J+1}=  e^{-U}
\left[ W(J)-W(J+1) \right]$ where we have used (\ref{avgfit}). 
For $s \to 0$, expanding the fitness $W(J)$ to leading orders in $s$, we get
\be
D_1(X_J) \approx A X_J \left(1- \frac{X_J}{{\cal X}_J} \right)
\label{D1}
\ee
where $A=c s \left[(J+1)^\alpha-J^{\alpha} \right] \approx \alpha c s
J^{\alpha-1}$ for large $J$.

The diffusion 
coefficient $D_2$ in (\ref{Dn}) gives the fluctuations in the frequency of the
least-loaded class about the mean value. These fluctuations arising due
to the finiteness of the population can be determined using
(\ref{binomial}) which gives the variance in the number of offspring
produced in one generation as \citep{Ewens:1979}
\be
D_2(X_J)= \frac{X_J (1-X_J)}{N} \approx \frac{X_J}{N}~.
\label{D2}~.
\ee
The last expression on the right hand side of the above equation
captures the fact that the fluctuations  vanish when either the population
size $N$ is infinite or the population is in the absorbing state
$X_J=0$.

Using the coefficients (\ref{D1}) and (\ref{D2}) in (\ref{TJ}), the
average inter-click time ${\overline T}_J$ can be written as 
\be
{\overline T}_J=2 N \int_0^{{\cal X}_J} dY ~ \psi(Y) \int_Y^1
\frac{dZ}{Z} ~\psi^{-1}(Z)
\label{TJ2}
\ee
where 
\be
\psi(Y)={\rm exp} \left[\beta \left( \frac{Y^2}{{\cal X}^2_J}- \frac{2 Y}{{\cal X}_J} \right) \right]
\ee
and 
\be
\beta= N {\cal X}_J A =N {\cal X}_J c s \left[(J+1)^\alpha-J^{\alpha}
  \right]
\label{beta}~.
\ee
In the absence of epistasis, both $A$ and ${\cal X}_J$ are independent
of $J$ so that typical 
time spent between any two successive clicks is constant and the
ratchet turns with a finite speed  
equal to $1/{\overline T}$. For epistatic
fitness $\alpha \neq 1$, ${\overline T}_J$ depends explicitly on $J$
and is expected to increase with $J$ for
$\alpha > 1$ while decrease for $\alpha < 1$
\citep{Kondrashov:1994}. In the following
discussion, we will restrict ourselves to $\alpha \geq 1$.

After some simple manipulations, we can rewrite (\ref{TJ2}) as 
\be
{\overline T}_J = 2 N {\cal X}_J \int_{-1}^0 dY e^{\beta Y^2}
\int_Y^{\delta} \frac{d Z}{1+Z} ~e^{-\beta Z^2}
\label{scaledTJ}
\ee
which implies that the scaled time ${\overline T}_J/(N {\cal X}_J)$ is
a function of two variables namely $\beta=N {\cal X}_J A$ and  
$\delta=(1-{\cal X}_J)/{\cal X}_J$. The 
nature of ${\overline T}_J$ depends on the parameter $\beta
\delta^2$ which can be seen as follows. Consider the Gaussian $e^{-\beta Z^2 }$  in the 
rightmost integral in (\ref{scaledTJ}) which is centred about $Z=0$
and has a width $1/\sqrt{\beta}$. If the
upper limit $\delta$ of this integral exceeds the width i.e. $\beta
\delta^2 \gg 1$, the integral
can be cutoff at $1/\sqrt{\beta}$ thus eliminating the dependence on
$\delta$. In such a case, ${\overline T}_J$ is of the following scaling form, 
\be
{\overline T}_J \approx N {\cal X}_J ~{\cal F}\left(N {\cal X}_J A
\right)
\label{scaling}
\ee
where ${\cal F}(\beta)$ is the scaling function determined below.  
If $\beta \delta^2 \ll 1$, then the inter-click time depends on
$\delta$ as well. Since
$\beta \delta^2 \sim N s J^{\alpha-1}/{\cal X}_J$ and ${\cal X}_J$ is
bounded above by one, $\beta \delta^2 \gg 1$ for large $J$ when
epistatic interactions are synergistic. For $\alpha=1$, the 
parameter $\beta \delta^2$ exceeds unity if $N s \gg 1$. Here we will
restrict ourselves to the $\beta \delta^2 \gg 1$ case which has nice
scaling properties although the double integral in
(\ref{scaledTJ}) can be estimated for $\beta \delta^2 \ll 1$ also.

We now proceed to find the scaling function ${\cal F}(\beta)$. If the
width of the Gaussian is large i.e. $\beta \ll 1$, we can approximate 
$e^{-\beta Z^2} \approx 1$ for $Z \ll
1/\sqrt{\beta}$ in the 
rightmost integral in (\ref{scaledTJ}). Since $\delta \gg
1/\sqrt{\beta}$, as argued above, this integral needs
to be carried out from $Y$ to $1/\sqrt{\beta}$. This yields 
\bea
{\overline T}_J &\approx& 2 N {\cal X}_J \int_0^1 dY e^{\beta Y^2}
\left[ \ln \left( 1+ \frac{1}{\sqrt{\beta}} \right)-\ln (1-Y) \right]
\no \\
&\approx& 2 N {\cal X}_J \left[1+ \ln \left(1+
  \frac{1}{\sqrt{\beta}} \right)\right] \no \\
&\approx& N {\cal X}_J \left(2-  \ln \beta \right)~,~
\beta \ll 1~.
\eea
To find the scaling function in the opposite limit $\beta \gg 1$, we
first consider the inner integral in (\ref{scaledTJ}), 
\bea
\int_Y^\delta \frac{dZ}{1+Z} ~e^{-\beta Z^2}&=& \frac{1}{\sqrt{\beta}}
\int_{Y \sqrt{\beta}}^{\delta \sqrt{\beta}}
dZ~e^{-Z^2} \left[1+ \frac{Z}{\sqrt{\beta}} \right]^{-1}  \no \\
&\approx& \frac{1}{2} \sqrt{\frac{\pi}{\beta}} \left[{\rm erf}
  \left(\delta \sqrt{\beta} \right)-{\rm erf}\left(Y \sqrt{\beta}
  \right) \right] 
 \eea 
where ${\rm erf}(z)$ is the error function \citep{Abramowitz:1964} and 
we have kept terms to leading orders in $\beta \delta^2 (\gg 1)$
and  $Y \sqrt{\beta}$ for $\beta \gg 1$. On using the above integral in
(\ref{scaledTJ}), we obtain
\bea
{\overline T}_J &\approx& \frac{N {\cal X}_J \sqrt{\pi}}{\beta}
\int_0^{\sqrt{\beta}} dY~ e^{Y^2} \left[ {\rm erf}
  \left(\delta \sqrt{\beta} \right)+{\rm erf}\left(Y \right)\right] \no
\\
&=& \frac{N {\cal X}_J \sqrt{\pi} e^{\beta}}{2 \beta} \int_0^{\beta}
\frac{dY e^{-Y}}{\sqrt{\beta-Y}}  \left[ {\rm erf}
  \left(\delta \sqrt{\beta} \right)+{\rm erf}\left(
  \sqrt{\beta-Y}\right)\right] \no \\
&\approx& N {\cal X}_J \sqrt{\frac{\pi}{\beta}} \frac{e^{\beta }}{\beta}
\eea
where we have used that both $\beta \delta^2$ and $\beta$ are large. 
From the above discussion, we find that the scaling function 
\be
{\cal F}(\beta) \approx
\begin{cases}
2-  \ln \beta~,~ \beta \ll 1 \\
 \sqrt{\pi} ~\beta^{-3/2} e^{\beta }~,~ \beta \gg 1
\end{cases}
\ee
is a ${\bf U}$-shaped function of $\beta$ reaching a minimum 
when $\beta \sim 1$.  We now discuss the above results in more
detail for $\alpha=1$ and $\alpha > 1$.


\noindent{\bf Non-epistatic fitness landscapes:} 
As we have already argued, when $\alpha=1$ the ratchet clicks with a
finite speed equal to $1/{\overline T}$.  When the rate at which
the ratchet clicks is large, analytical results can 
be obtained using the traveling wave approach
\citep{Rouzine:2003}. For slowly clicking ratchet which is the subject
of this article, the problem was formulated analytically within a 
diffusion approximation first by \citet{Stephan:1993}. However a
better agreement between the diffusion theory and the simulation
results was obtained in 
\citet{Gordo:2000a,Gordo:2000b}. A possible reason for this difference is that 
\citet{Stephan:1993} included terms besides those in (\ref{D1new}) 
in the expansion of the drift coefficient which is not consistent with the
assumption of linear response, while the expression for $D_1$ in
\citet{Gordo:2000a} is the same as (\ref{D1new}). 
In fact, the
expression for the inter-click time given by (\ref{scaledTJ}) with
$c=0.6$ is identical
to that reported in \citet{Gordo:2000a}. However 
the integrals were computed numerically by these authors 
while here we estimate them analytically and find that 
the average inter-click time is given by  
\be
{\overline T} \approx 
\begin{cases}
n_0 \left(2- \ln \beta \right)~,~\beta \ll 1 \\
\sqrt{\pi} n_0 ~\beta^{-3/2} e^{\beta}~,~\beta \gg 1 
\end{cases}
\label{Tmult}
\ee
where $n_0=N {\cal X}_J=N e^{-U/s}$ is the number of individuals in
the least-loaded class and $\beta=n_0 c s$ is the scaling
parameter. The inter-click time calculated numerically using the integral
(\ref{scaledTJ}) and the above expression (\ref{Tmult}) is shown in
Table \ref{mult} and the two are seen to be in good agreement.

\begin{figure}
\includegraphics[width=0.6 \linewidth,angle=270]{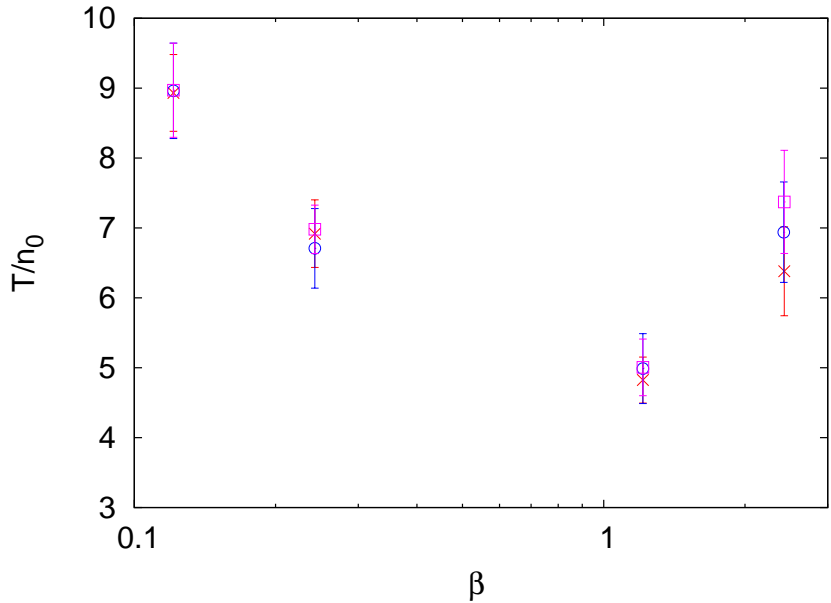}
\caption{Data collapse of the scaled time ${\overline T}/n_0$ when
  plotted against
  the scaling parameter $\beta=n_0 c s$ for the multiplicative
  fitness landscape where $c=0.6$. The simulation data has been
  averaged over $100$ stochastic runs. }
\label{collapse}
\end{figure}

Our first result concerning the Muller's ratchet is the scaling form for
time ${\overline T}$ when parameters $N, U$ and $s$ are chosen 
such that $\beta \delta^2 \sim N s \gg 1$ is satisfied. The results of our
numerical simulations testing this scaling form 
are shown in Fig.~\ref{collapse} where we have
differentiated data points at fixed $\beta$ for clarity. The scaled
time ${\overline T}/n_0$ indeed shows a very good
data collapse and a non-monotonic dependence on $\beta$. 
\citet{Higgs:1995} unsuccessfully 
attempted to obtain a scaling form for the ratchet rate by using $N s$ as
the scaling parameter for $N s \gg 1$. The scaling function in (\ref{Tmult})
 however shows that the scaling parameter is a function of the population
 number $n_0$ in the least-loaded class, and not the total population $N$. 

\begin{table}
\begin{center}
\begin{tabular*}{0.9\textwidth}{@{\extracolsep{\fill}}llllrlll}
\hline
$\beta$ & $N$ & $U$ & $s$ & $n_0$ & ${\overline T}_{\rm sim}$(2 SE) &
${\overline T}_{\rm int}$ &
${\overline T}_{sca}$  \\
\hline
0.121 & 7500 &	0.02 &	0.004 &	50.5 &	453(34) & 230 & 207\\
  &	15000 &	0.01 &	0.002 &	101.1 &	906(69) & 460 & 416\\
  &	30000&	0.005 &	0.001 &	202.1 &	1805(111) & 920 & 831 \\ 
\hline
0.242 & 3000 &	0.1 &	0.02 &	20.2 &	141(7) & 86 & 69\\
  &	6000 &	0.05 &	0.01 &	40.4 &	271(23) & 172 & 138\\
  &	12000 &	0.025 &	0.005 &	80.8 &	559(39) & 344 & 276\\
\hline
1.212 & 15000 &	0.1 &	0.02 &	101.1 &	506(41) & 479 & 451\\
  &	30000&	0.05 &	0.01 &	202.2 &	1008(101) & 959 & 902\\
  &	60000&	0.025 &	0.005 &	404.2 &	1950(133) & 1918 & 1804\\
\hline
2.424 & 30000 &	0.1 &	0.02 &	202.1 &	1490(149) & 1487 & 1071 \\
  &	45000 &	0.067  & 0.013 & 303.2 & 2104(218) & 2230 & 1608\\
  &	60000 &	0.05 &	0.01 &	404.2 &	2579(257) & 2974 & 2143\\
\hline
\end{tabular*}
\caption{Comparison of simulation data plotted in Fig.~\ref{collapse}
  with analytical results. Here
  ${\overline T}_{\rm sim}$ refers to the inter-click time obtained in
  simulations, ${\overline T}_{\rm int}$ by numerically evaluating the integrals
  in (\ref{scaledTJ}) for $\alpha=1$ and ${\overline T}_{sca}$ using
  the scaling form (\ref{Tmult}).}
\label{mult}
\end{center}
\end{table}

In many studies \citep{Haigh:1978,Bell:1988,Gessler:1995}, the size
$n_0$ of the least-loaded class has been regarded as an important parameter
in determining the ratchet speed. If $n_0 \gg 1$, the population is
close to the deterministic limit and the ratchet clicks slowly whereas for
$n_0 \ll 1$, the ratchet speed is high. However 
the simulations show that the size $n_0$ of the least-loaded class is
not sufficient to predict the ratchet rate
\citep{Stephan:1993,Gordo:2000a}.  This is indeed captured by
diffusion approximation as (\ref{Tmult}) is not a function of
$n_0$ alone.  However, if $s$ is kept fixed,  ${\overline T}$ increases
monotonically with $n_0$ in accordance with the above expectation and 
simulations \citep{Gabriel:1993,Gordo:2000a}. The dependence of ${\overline T}$ 
on $s$ for given $n_0$ is however
non-monotonic similar to that seen in numerical studies
\citep{Gordo:2000a}. To understand this behavior qualitatively,
consider the situation where the population number $n_0$ is kept constant 
by keeping $N$ and $U/s$ fixed. As increasing $s$ tends to 
 localise population and hence increase ${\overline T}$, 
increasing $U$ has the opposite delocalising effect which decreases 
${\overline T}$. At a given $U$ and $s$, one of these two competing
forces wins. According to (\ref{Tmult}), as the
scaling function overturns when $n_0 s \sim 1$, 
the mutation takes over for $U > U^*=s \ln (N s)$ whereas below $U^*$, 
selection dominates and ${\overline T}$ is large.

The solution (\ref{Tmult}) gives an
initial logarithmically slow drop in $s$ and an exponential increase for larger
$s$ with the minimum of the ${\bf U}$-shaped curve 
occuring at a selection coefficient which scales as $1/n_0$. The
simulation results of \citet{Gordo:2000a} (also see Table \ref{mult}) 
however show a much faster
drop at small $s$. A good agreement with simulation data was 
obtained in \citet{Gordo:2000b} 
by adding ${\overline  T}$ and time $T_a \sim 1/s$
required by the population to relax to new steady state just after a click. 
 However, the full expression for $T_a$ is not of scaling form
 (\ref{scaling}) although  
the simulation data in Fig.~\ref{collapse} shows an
excellent data collapse even for small $s$. In view of this, a better
understanding of the time $T_a$ is desirable. 
Of course, for both $n_0$ and $s$  fixed, the time ${\overline T}$ is
predicted to be independent of $N$ and $U$ which is confirmed by simulations 
\citep{Gordo:2000a}.

\begin{figure}
\centerline{\includegraphics[width=0.42 \linewidth,angle=270]{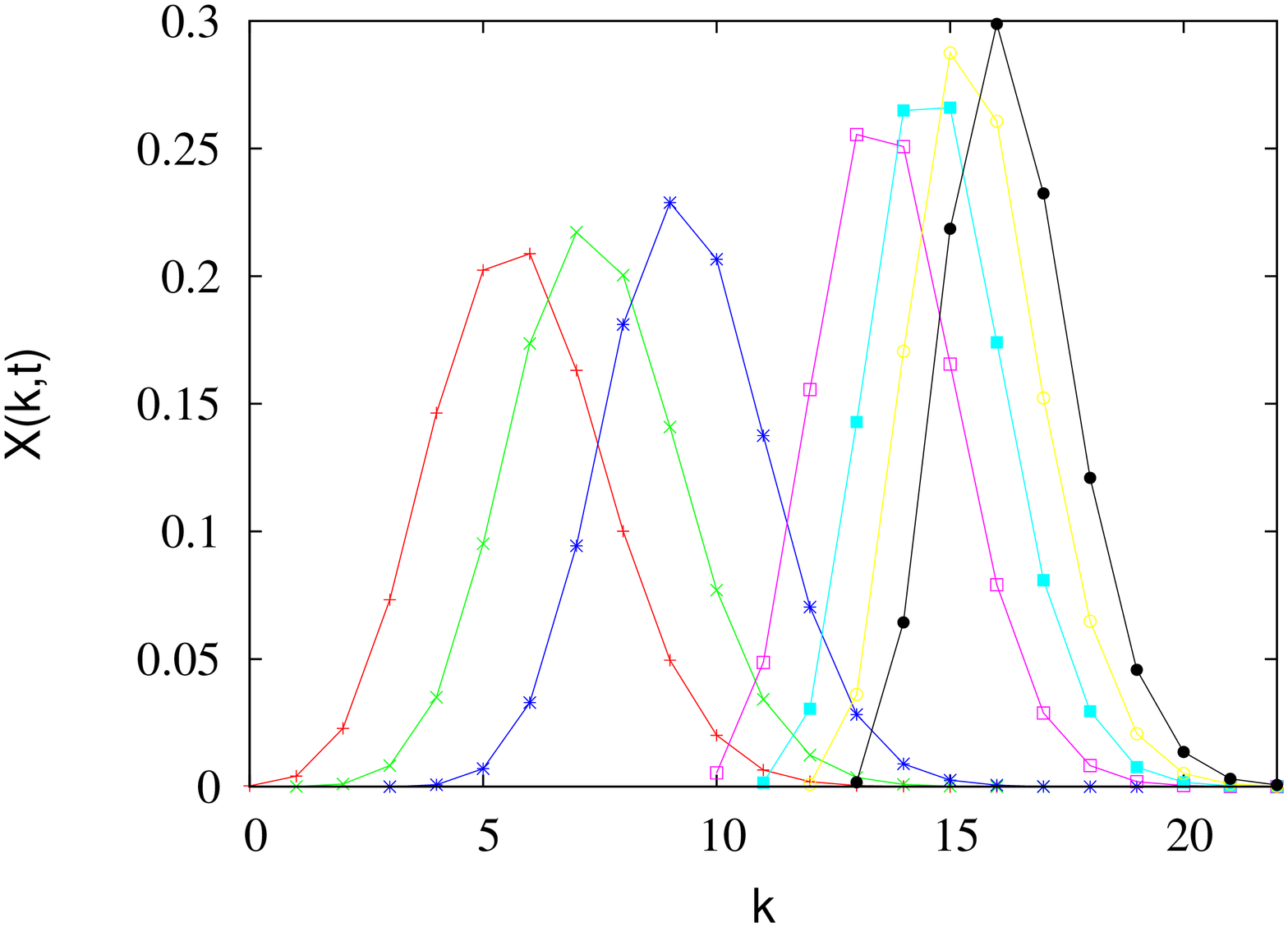}
\includegraphics[width=0.42 \linewidth,angle=270]{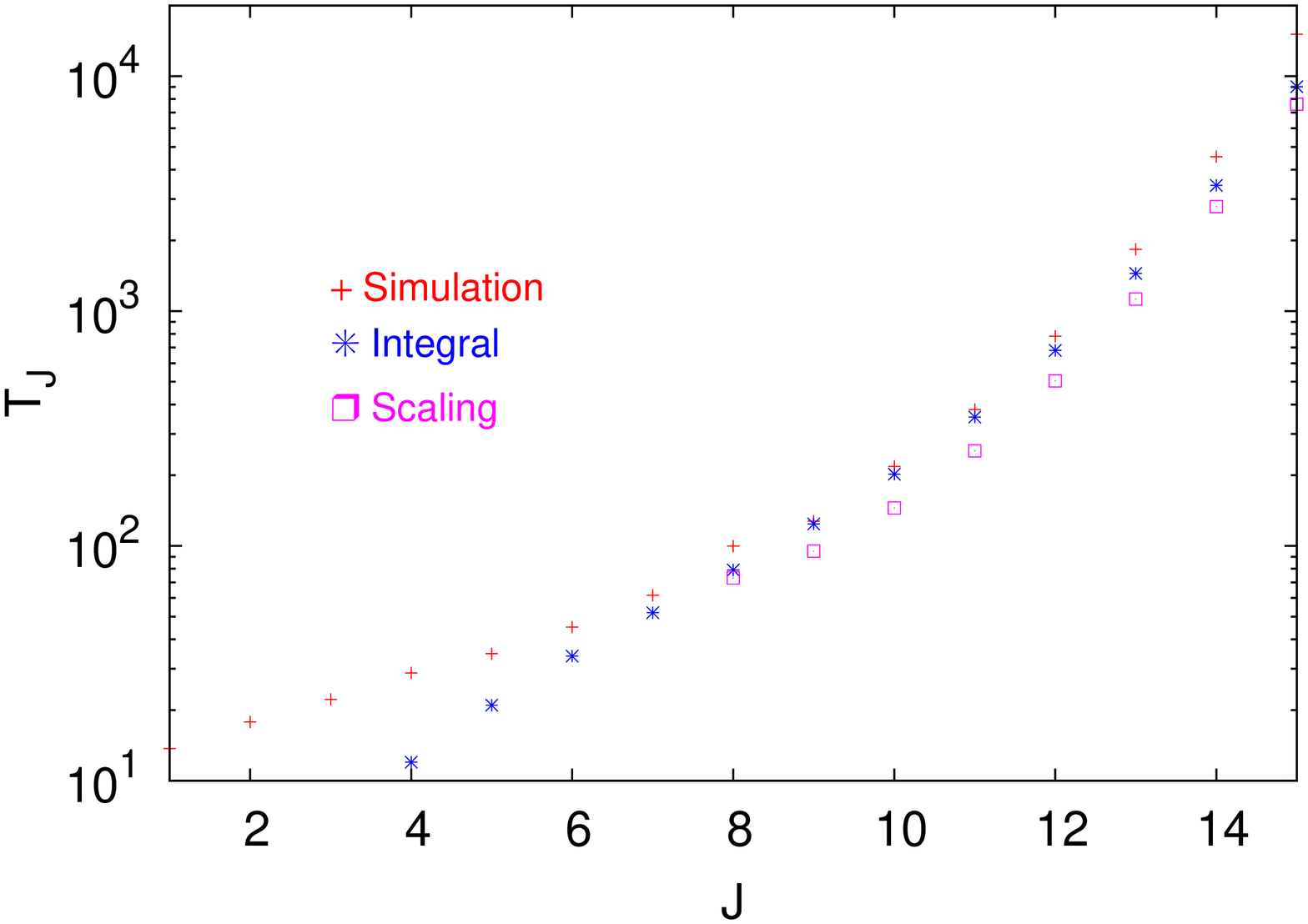}}
\caption{Loss of least-loaded class on epistatic fitness landscapes with
$\alpha=2, N=256, U=0.25, s=0.005$: (a) Average fraction ${\overline
    X}(k,t)$ in the $k$th class at $t=50, 100, 200, 1250, 2500, 5000,
10000$ (from left to right). The data have been averaged over 
$1000$ stochastic histories. (b)  Average inter-click time ${\overline
  T}_J$ as a function of $J$ where each data point has been averaged over $500$
  histories. Comparison with ${\overline T}_J$ obtained by evaluating integral
(\ref{scaledTJ}) numerically and using the scaling
  form (\ref{scaling}) at large $J$ is also
  shown. A reasonably good agreement is obtained with $c=0.9$.}
\label{epi}
\end{figure}

\noindent{\bf Epistatic fitness landscapes:} For synergistic
interactions, the ratchet is expected to halt at large times  
\citep{Charlesworth:1993,Kondrashov:1994}. Figure \ref{epi}a shows the 
population 
fraction $\overline{X}(k,t)$ with $k$ deleterious mutations at several
time slices. Two points are noteworthy: the
ratchet does not turn with a constant speed as is evident by the rate at which the
population accumulates average number of
mutations at late times. 
 Secondly, unlike for the multiplicative
fitness case \citep{Rouzine:2003}, the population fraction does not
maintain its shape as the width of the
distribution $\overline{X}(k,t)$ decreases with increasing time. Thus 
the number frequency of  
an asexual population under epistatic selection does not behave like
a traveling wave moving with a constant speed.

The simulation data for average inter-click time shown in
Fig.~\ref{epi}b increases with the minimum number $J$ of deleterious
mutations in the population thus indicating the arrest of the ratchet
at large times. The time ${\overline T}_J$ obtained by integrating
(\ref{scaledTJ}) for $\alpha=2$ is also shown for comparison and we
find that it agrees well with the simulation results. 
As the scaling parameter $\beta$ defined in (\ref{beta})
increases as a power law with $J$ for $J \gg 1$, we have $\beta \gg 1$
for large $J$. In such a case, ${\overline T}_J \sim N {\cal X}_J
\beta^{-3/2} e^{\beta}$ which increases exponentially fast with
$\beta$. As ${\cal X}_J \to 1$ for large $J$, the inter-click time ${\overline T}_J \sim e^{N
\alpha c s J^{\alpha-1}}$ increases faster than any power law with $J$
and for any $\alpha > 1$. Thus an arbitrarily small $\alpha-1$ is
capable of slowing down the ratchet under synergistic selection.

To estimate how the ratchet speed approaches zero, we use the following
argument. The average speed ${\overline v}=d/t$ where $d$ is the
number of minimum mutations accumulated in time interval $t$. Since 
$T_J$ is the time between the $(J-1)$th and $J$th click of the ratchet, we have
\be
{\overline v} =  \overline{\frac{d}{T_0+...+T_d}}
\ee
where $d$ is fixed. Assuming the distribution for click times has nice
scaling properties, we may write
\be
{\overline v} \sim \frac{d}{\overline{T_0+...+T_d}}~.
\ee
If $T_J$ does not depend on the least-loaded class $J$, we have
${\overline v}= 1/{\overline T}$ as expected for $\alpha=1$ case. For
$\alpha > 1$ as the average
inter-click time increases faster than power law with $J$, the sum 
$\overline{T_0+...+T_d} \sim d^{2-\alpha} e^{N
\alpha c s d^{\alpha-1}}$ which is reasonable as
the sum is   dominated by $\overline{T_d}$ for large $d$. This gives 
${\overline v} \sim d^{\alpha-1} e^{-N
\alpha c s d^{\alpha-1}}$ or in terms of $t$, 
\be
{\overline v} \sim \frac{1}{t}~\left( \frac{\ln t}{N \alpha c s} \right)^{1/(\alpha-1)}~.
\ee
Thus in the presence of epistasis, the average speed of the ratchet 
approaches zero as $1/t$ with $\alpha$-dependent logarithmic corrections.

\bigskip
\centerline{DISCUSSION}
\bigskip

In this article, we considered the effect of drift and epistasis on
the loss of the least-loaded (or the fittest) class in an asexual 
population. When the population size is
infinite, the drift is absent and the population evolves due to the 
elementary processes of selection and mutation. As selection tends to
localise the population at the fittest sequence while mutation has the
opposite tendency to delocalise it, an error threshold may exist 
beyond which the fittest class cannot be sustained in the 
population \citep{Eigen:1971}. Such a phase transition is known to
occur for asexual
populations evolving deterministically on fitness landscapes defined 
by (\ref{fitness}) when $\alpha < 1$ \citep{Wiehe:1997}. However for $\alpha
\ge 1$, the population frequency ${\cal X}_J$ in the least-loaded
class $J$ remains nonzero for any finite $U, s$. In fact, for synergistic
interactions, the frequency ${\cal X}_J$ 
given by (\ref{synXJ}) increases with $J$ towards unity. 

If however the population size is finite, as illustrated in
Fig.~\ref{synrun}, the population frequency
$X_J(t)$ fluctuates with time and can become zero even for $\alpha \ge
1$. Numerical simulations have shown that this loss occurs at a
constant speed when $\alpha=1$ 
\citep{Haigh:1978} but at a decelerating rate when
$\alpha > 1$ \citep{Charlesworth:1993,Kondrashov:1994}. 
In this paper, we focused on the stochastic dynamics of the 
loss of the fittest class for $\alpha \geq 1$ 
and calculated
explicit analytical expressions for the typical time ${\overline T}_J$
during which the least-loaded class $J$ survives using a diffusion theory
\citep{Ewens:1979}. Although this approach has been considered
previously to attack the Muller's ratchet problem
\citep{Stephan:1993}, the resulting
solution in the form of a double integral was evaluated numerically 
which does not allow one to infer the functional dependence of
${\overline T}_J$ on parameters $N, U, s$ and $J$.

When the interactions between gene loci are assumed to be absent and
$N s \gg 1$, the inter-click time ${\overline T}$ is found to be 
of the scaling form (\ref{Tmult}). 
Although this result is derived using 
diffusion theory which is based on several approximations, the numerical
simulations show an excellent data 
collapse suggesting that (\ref{scaling}) may be an exact statement. 
For fixed $n_0$, the time ${\overline T}$ is seen to be 
a ${\bf U}$-shaped function of $s$ arising due to competition between
mutation and selection. 
Such a behavior is reminiscent of error threshold
phenomenon in infinite populations discussed above. Although the least-loaded
class is never lost in the deterministic limit on multiplicative
fitness landscapes \citep{Wagner:1993}, the selection-mutation
competition manifests itself in the time 
duration during which the least-loaded (fittest) class can support a
finite population. For given $s$, the survival time ${\overline T}$
initially increases linearly with
$n_0$ approaching the deterministic limit of $N \to \infty$ with an
exponential rise with $n_0$ as increasing $N$ decreases the effect of
drift.

Muller's ratchet has been proposed as a possible mechanism for the
degeneration and eventual extinction of asexual organisms and
nonrecombining parts of sexually reproducing populations. For
bacterial populations with $N \sim 10^6$, $U \approx 0.003$
\citep{Drake:1998} and $s \approx 10^{-3}$, using (\ref{Tmult}) we find
it takes $10^{15}$ generations for one click to occur which does not
seem plausible. However, reducing $s$ by a factor half gives
${\overline T} \sim  5000$ and further reduction to $s \approx 0.25
\times 10^{-3}$ gives just $50$ generations for a single
click. Muller's ratchet has also been invoked to understand the 
degeneration of the nonrecombining neo-Y chromosome which originated 
about a million years ago in {\it  Drosophila Miranda}
\citep{Charlesworth:1978,Gordo:2000a}. Assuming that
a few thousand deleterious mutations have occurred over this time
span,  the time between successive turns of the ratchet is of the order of
$10^3$ generations. From (\ref{Tmult}), the time ${\overline T} \sim
e^{n_0 c s}$ for large ${\overline T}$ which gives $n_0 s \approx
14$. If $s \approx 10^{-2}$, the population size $N$ required for the
Muller's ratchet to be a viable mechanism works out to be $\sim
700 e^{100 U}$ which depends sensitively on $U$ due to the exponential
dependence. For instance, for $U=0.07$, the required population is of the order
$5 \times 10^5$ while it reduces by a factor twenty for $U=0.04$. A
similar sensitive dependence of extinction time on $s$ for given
$N$ and $U$ has been 
noted in the problem of the degeneration of human mitochondrial DNA
also \citep{Loewe:2006}.  This
suggests that very precise estimates of $U$ and $s$ may be 
required to determine whether the Muller's ratchet might be in
operation. 

The scaling form (\ref{scaling}) holds for
$\alpha > 1$ also but the speed of the ratchet under epistatic
selection is found to decay rapidly with time for any $\alpha > 1$. 
Although the Muller's ratchet under synergistically epistatic
selection has the interesting feature of arresting the loss of
least-loaded class, the generality of this mechanism seems unclear 
as general support for synergistic epistasis has not been
found in the experiments \citep{Visser:2007} and the slowing down
effect is sensitive to the
inclusion of biologically relevant details such as 
distribution of mutational effects \citep{Butcher:1995}.
Recent experimental
evidence suggests that fitness decline down to a
plateau  can be attributed to the presence of
epistasis \citep{Silander:2007}.  This can be due to negative epistasis 
\citep{Kondrashov:1994} or compensatory epistasis
\citep{Silander:2007}. It would be interesting to 
compare the results discussed in this work with models that include 
compensatory mutations \citep{Wagner:1990,Wilke:2003}.

Acknowledgement: The author is grateful to J. Krug, S.-C. Park and an
anonymous refereee for helpful comments on the manuscript.


\end{document}